\documentclass{article}

\usepackage{amssymb} 
\usepackage{amsmath} 
\usepackage{latexsym} 

\usepackage[dvips]{psfrag,graphicx} 

\numberwithin{equation}{section}

\def\G{\Gamma}
\def\RR{\mathbb{R}}
\def\e{\varepsilon}
\def\vect#1{\mbox{\boldmath $#1$}} 

\begin{document}

\title{Ideal, best packing, and energy minimizing double helices 
}

\author{Jun \textsc{O'Hara}}

\maketitle

\begin{abstract}
{
We study optimal double helices with straight axes (or the fattest tubes around them) computationally using three kinds of functionals; ideal ones using ropelength, best volume packing ones, and energy minimizers using two one-parameter families of interaction energies between two strands of types $r^{-\alpha}$ and $\frac1r\exp(-kr)$. We compare the numerical results with experimental data of DNA. 
}
\end{abstract}



\maketitle

\section{Introduction}
In this paper\footnote{to appear in Progress of Theoretical Physics Supplement, proceedings of ``Statistical physics and topology of polymers with ramifications to structure and function of DNA and proteins'', Kyoto 2010.} we study three kinds of functionals for a one-parameter family of infinitely long double helices with straight axes that have radius $1$ and {\em slope} $a>0$, i.e. {\em pitch} $P=2\pi a$: 
$$
\begin{array}{c}
\G(a)=\G_1(a)\cup \G_2(a) \>\>(a>0),\\[1mm]
\G_1(a)=\{(\cos\theta, \sin\theta, a\theta)\}, \>\G_2(a)=\{(-\cos\theta, -\sin\theta, a\theta)\},
\end{array}
$$
%
and compute the optimal slopes with respect to them numerically. 

The first functional is the ``{\sl average ropelength}'' which is the smallest length of a rope with unit thickness which is needed to make one twist of the double helix. It is a dual notion of {\em thickness}\cite{KV} which is the radius of the ``fattest tube'' around a curve, or equivalently, the minimum {\em global radius of curvature}\cite{Go-Ma}. To be precise, the thickness of a curve $\G$, which we denote by $\rho(\G)$, is given by the supremum of the radii $r>0$ so that a tubular $r$-neighbourhood of $\G$ (denoted by $N_r(\G)$) can be embedded, i.e. the supremum of $r>0$ so that the normal discs to $\G$ of radius $r$ at distinct points never intersect \cite{LSDR,Bu-Si3,Ku-Su2}. 

The {\em ropelength} is the ratio of the length $L(\G)$ of $\G$ and the thickness $\rho(\G)$\cite{Ku-Su1',LSDR}. A knot that minimizes the ropelength in its isotopy class is called an {\em ideal knot} after Stasiak. In other words, an ideal knot is a knot whose fattest tubular neighbourhood maximizes a scale invariant functional on the space of knots defined by 
$$K\mapsto \frac{\textrm{Vol}(N_{\rho(K)}(K))}{L^3(K)}$$
in its isotopy class. 
The ideal double helix was computed by Sylwester Przybyl and Piotr Piera\'nski\cite{PP,P}. 

The second is the ``{\sl packing proportion}'', in other words, the ratio of the volume of the fattest tubes and that of the circumscribed cylinder of the tubes (Figure \ref{0635805_cylinder}). The best volume packing double helix was computed by Kasper Olsen and Jakob Bohr\cite{OB}.

The last is an ``{\sl average mutual energy}'', which is a central topic of the paper. An {\em energy of knots} is a functional on the space of knots which was introduced to produce a representative embedding for each isotopy class as an energy minimizer in the isotopy class. 
The first example was obtained as the renormalization of $r^{-2}$-modified potential energy of knots\cite{O}. This energy is sometimes called the {\em M\"obius energy} because it is invariant under M\"obius transformations of the ambient space, but in fact, there are many other energies which are also invariant under M\"obius transformations\cite{LO,OS}. The energy blows up if a knot degenerates to a singular knot with double points. Therefore a knot can be deformed to decrease its energy without having self-crossing, and hence one can expect that the isotopy class would be kept the same during the deformation. 

Many kinds of energies have been studied intensively for about twenty years, forming a branch of mathematics called ``physical knot theory''. In this paper we use two  one parameter families of energies for $\G_1(a)\cup\G_2(a)$, 
$$
\int_{\G_1(a)}\int_{\G_2(a)}r^{-\alpha}\,dxdy \>(\alpha>1)\>\>\mbox{ and }\>\>
\int_{\G_1(a)}\int_{\G_2(a)}\frac1r\exp(-kr)\,dxdy \>(k>0)\,,
$$
where $r$ denotes the distance $|x-y|$ between a pair of points. Namely, we consider the integration of modified Coulomb's potential and the screened Coulomb's potential (Yukawa potential). 
As both diverges because of non-compactness of the domain of integration, we take the ``average per one twist'', in other words, we consider the integration on $I_1(a)\times\G_2(a)$, where $I_1(a)$ is a subarc of $\G_1(a)$ that winds up just once around the axis. As we only consider the interaction between two strands $\G_1(a)$ and $\G_2(a)$, the energies are well-defined without renormalization. We compute the slopes that minimize the above energies and see how they depend on the parameters $\alpha$ and $k$. 

We will compare our numerical results with experimental data of DNA reported by Stasiak and Maddocks. 

We remark that our strategy of looking for optima double helices with respect to one of the three kinds of functionals above mentioned does not work for single helices. In fact, our functionals are all optimized when the slope $a$ goes to $+\infty$, i.e. when the curve approaches a straight line.

\section{An ``ideal'' double helix}
\subsection{Thickness}
Let $\G$ be a curve. 
Let $d:\G\times \G\to\RR$ be the distance function: $d(x,y)=|x-y|$. 
The {\em doubly critical self distance} of $\G$\cite{Si3,LSDR}, denoted by $\textrm{dcsd}(\G)$, is the smallest positive critical value of $d$, i.e. 
$$
\textrm{dcsd}(\G)=\min_{x\ne y}\{|x-y|\,:\,\vect{t}_x, \vect{t}_y\perp(y-x)\},
$$
where $\vect{t}_x$ denotes the tangent vector to $\G$ at $x$. Then the thickness of a knot $\G$ is given by \cite{LSDR}
\begin{equation}\label{f_thickness}
\rho(\G)
=\displaystyle \min\left\{\min_x\{\mbox{radius of curvature at $x$}\}, \,\frac12\textrm{dcsd}(\G)\right\}.
\end{equation}
Remark that, for a fixed point $x_0$, the radius of curvature at $x_0$ is a local quantity whereas $\min_{y\ne x_0}\{|x_0-y|\,:\,\vect{t}_{x_0}, \vect{t}_y\perp(y-x_0)\}$ is a global quantity. 

We remark that, in the case of a {\sl single} helix with slope $a$, Maritan, Micheletti, Trovato, and  Banavar\cite{MMTB} showed that there is a critical slope $a_0\approx0.399805\approx2.512/2\pi$ such that the thickness is given by the half of the doubly critical self distance if $a<a_0$ and by the minimum radius of curvature if $a\ge a_0$. 


\subsection{Average ropelength and the slope of an ideal double helix}
%
Let $AL(a)$ be the length of $\G(a)$ per one twist. 
It is equal to $4\pi\sqrt{1+a^2}$. 
Define the {\em average ropelength} of a double helix $\G(a)$ with slope $a$, denoted by ${ARL}(a)$, by 
$$
{ARL}(a)=\frac{AL(a)}{\rho(a)}=\frac{4\pi\sqrt{1+a^2}}{\rho(a)},
$$ 
where $\rho(a)$ denotes the thickness of $\G(a)$. Let us compute the slope $a$ of an {\em ideal double helix}, i.e. the slope $a$ that attains the minimum average ropelength. 

The thickness $\rho(a)$ of $\G(a)$ is given as follows. 
The radius of curvature is constantly equal to $1+a^2$. On the other hand, the doubly critical self distance is equal to the distance between a point $(1,0,0)\in \G_1(a)$ and $\G_2(a)$. The latter is given by $2$ if $a\ge1$ and by 
$$\sqrt{(\cos\theta_0+1)^2+\sin^2\theta_0+a^2{\theta_0}^2}=\sqrt{2+2\cos\theta_0+a^2{\theta_0}^2},$$
where $\theta_0$ $(0<\theta_0<\pi)$ is given by $\sin\theta_0=a^2\theta_0$ if $a<1$. 
Therefore, by \eqref{f_thickness}, the thickness is given by the half of the doubly critical self distance for any $a$: 
\begin{equation}\label{rho}
\rho(a)=\left\{\begin{array}{cll}
\>\>1 && \>\>(a\ge1)\\[1mm]
\frac12\sqrt{2+2\cos\theta_0+a^2{\theta_0}^2}\>,&\> (0<\theta_0<\pi, \sin\theta_0=a^2\theta_0) & \>\>(0<a<1).
\end{array}
\right.
\end{equation}

\begin{figure}[htbp]
\begin{center}
\begin{minipage}{.45\linewidth}
\begin{center}
\includegraphics[width=\linewidth]{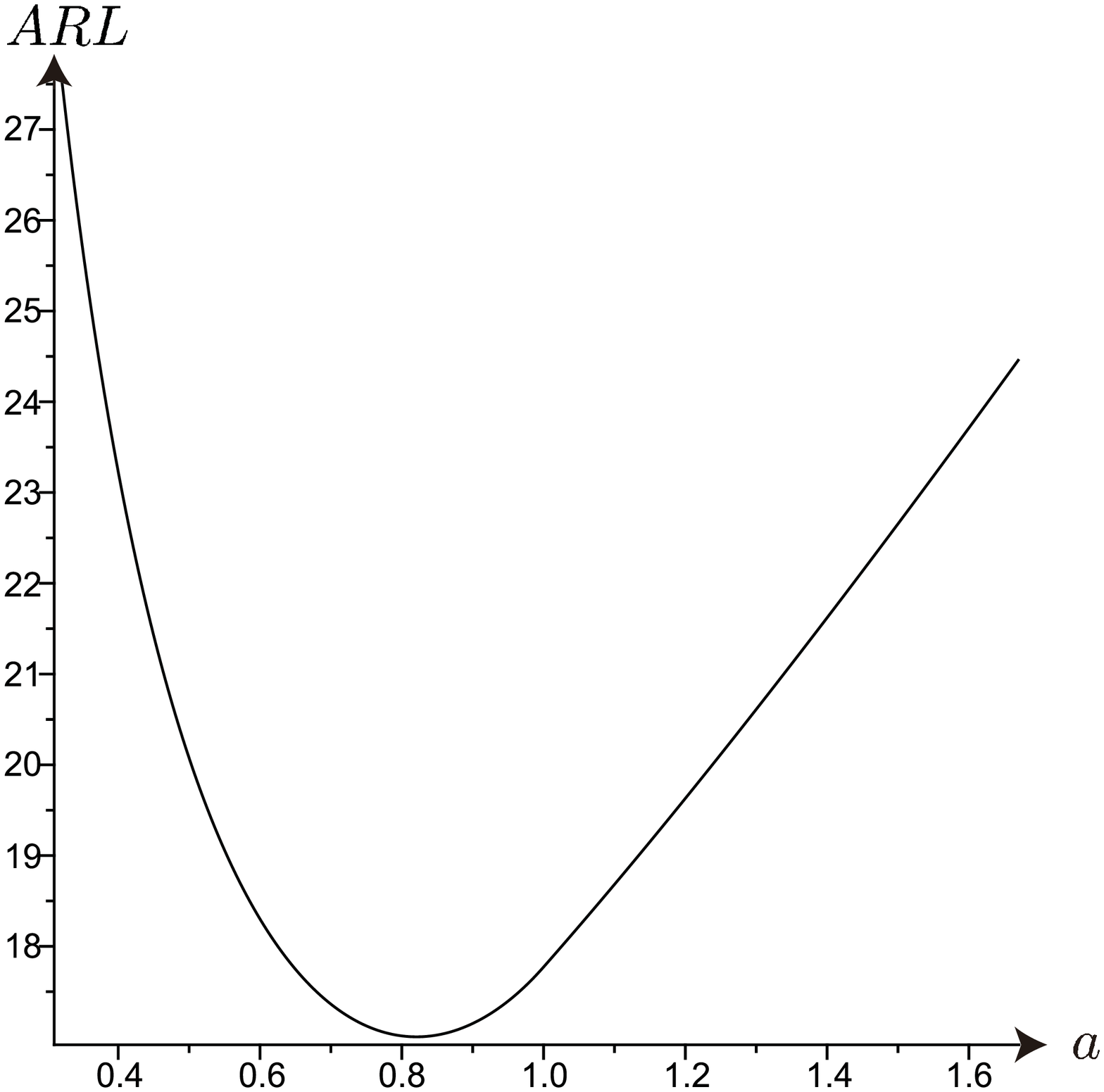}
\caption{The average ropelength of a double helix with slope $a$ }
\label{av_t_g_c2}
\end{center}
\end{minipage}
\hskip 0.2cm
\begin{minipage}{.5\linewidth}
\begin{center}
\includegraphics[width=0.8\linewidth]{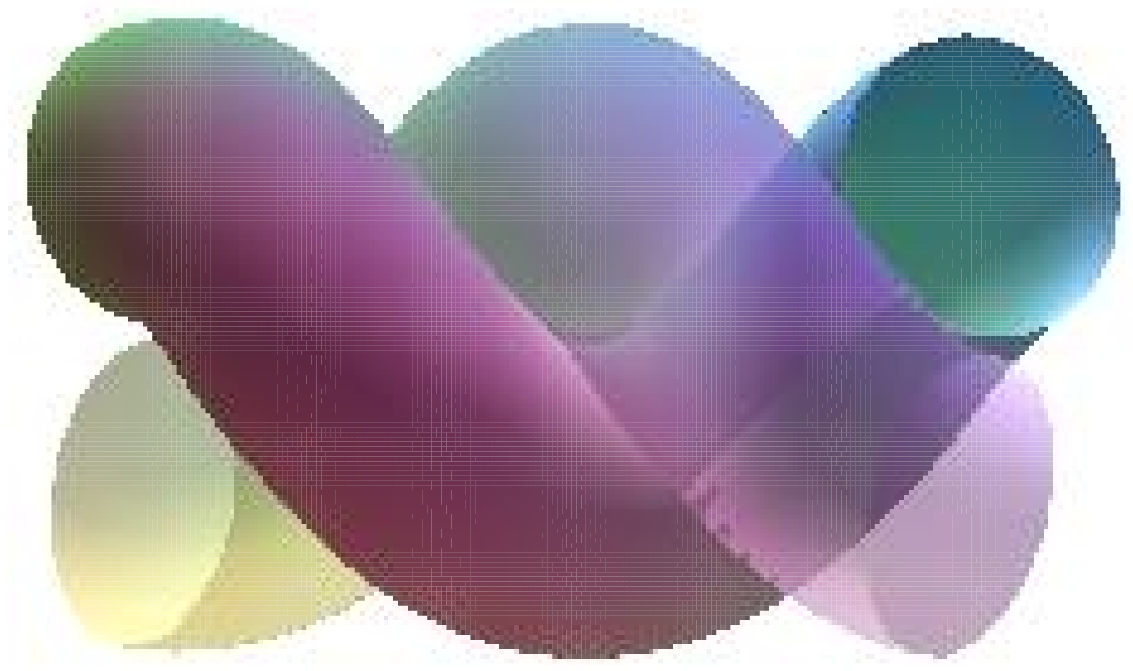}
\caption{Optimal tubular double helix with respect to the ropelength. This ideal double helix had already been reported by Pieranski\cite{P}{}, Fig. 12 (b). }
\label{082074_wogrid}
\end{center}
\end{minipage}
\end{center}
\end{figure}

When $a\ge1$ we have $ARL(a)=4\pi\sqrt{1+a^2}$ which takes the minimum value $4\sqrt{2}\pi\approx17.7715$ at $a=1$. The numerical computation using Maple implies that the average ropelength 
takes the minimum which is approximately equal to $17.003$ at $a\approx 0.82074$ (Figure \ref{av_t_g_c2}), when the thickness $\rho$ is $0.95614$ and the ratio of the pitch and the thickness is $P/\rho=2\pi a/\rho(a)\approx5.3934$. 

The above result is not new: 
This double helix with a straight axis that is optimal with respect to the ropelength (Figure \ref{082074_wogrid}) had already been reported by Pieranski\cite{P}{}, Fig. 12 (b). 

The equation \eqref{rho} implies that the closest point in $\G_2(a)$ from a point 
in $\G_1(a)$ can be found in the opposite side when $a\ge1$ and in an upward (or downward) direction  when $a<1$. 
It means that the curve of the contact of two fattest tubes around $\G_1(a)$ and $\G_2(a)$ is the straight axis when $a\ge1$ i.e. when the ratio of pitch and thickness satisfies $P/\rho\ge2\pi$, and a helix between $\G_1(a)$ and $\G_2(a)$ when $0<a<1$ i.e. when $P/\rho<2\pi$, as was explained by Stasiak and Maddocks\cite{SM}.

\section{Best volume packing double helix}
\begin{figure}[htbp]
\begin{center}
\begin{minipage}{.55\linewidth}
\begin{center}
\includegraphics[width=.75\linewidth]{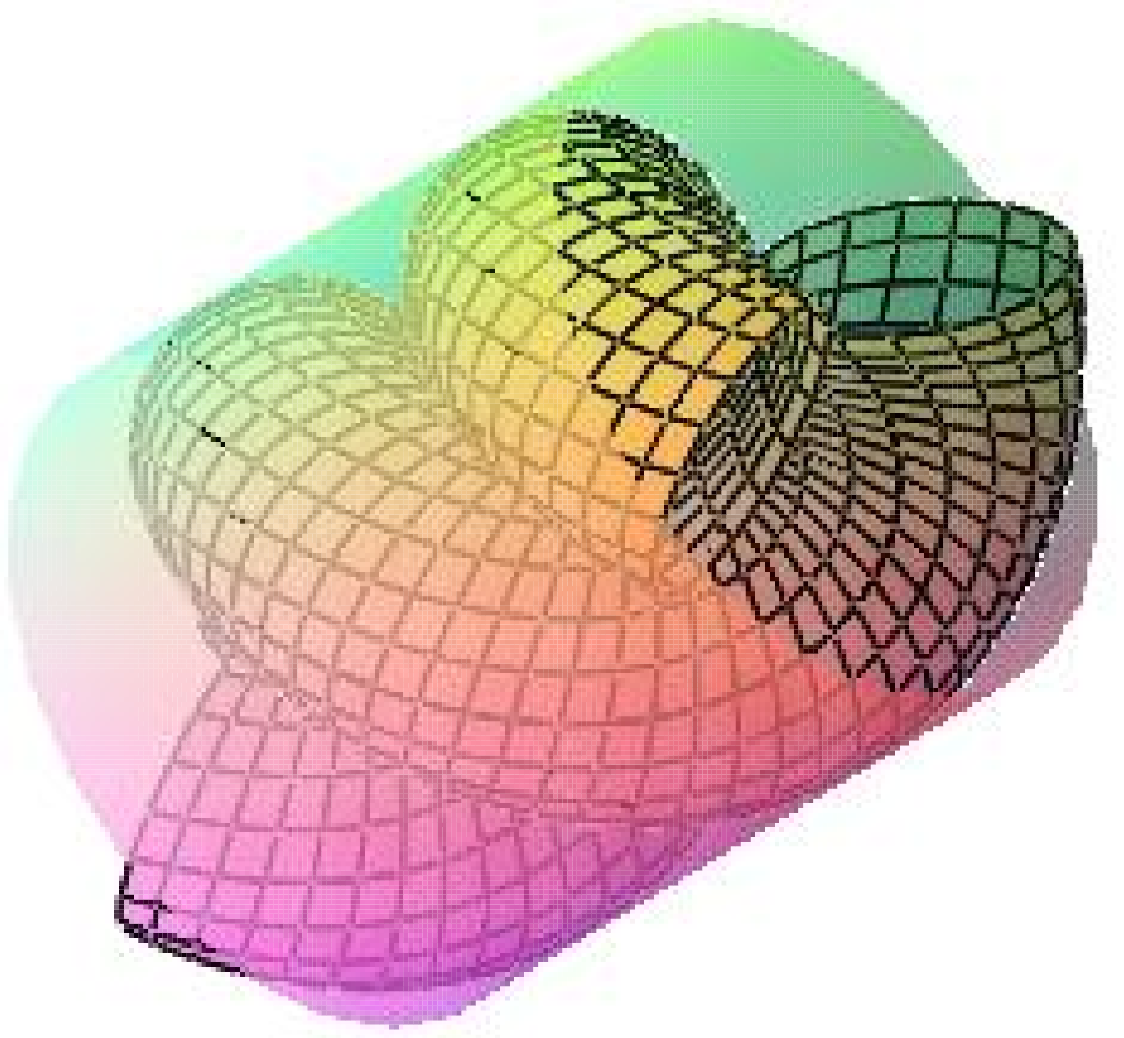}
\caption{Best volume packing tubular double helix and the circumscribed cylinder}
\label{0635805_cylinder}
\end{center}
\end{minipage}
\hskip 0.4cm
\begin{minipage}{.4\linewidth}
\begin{center}
\includegraphics[width=\linewidth]{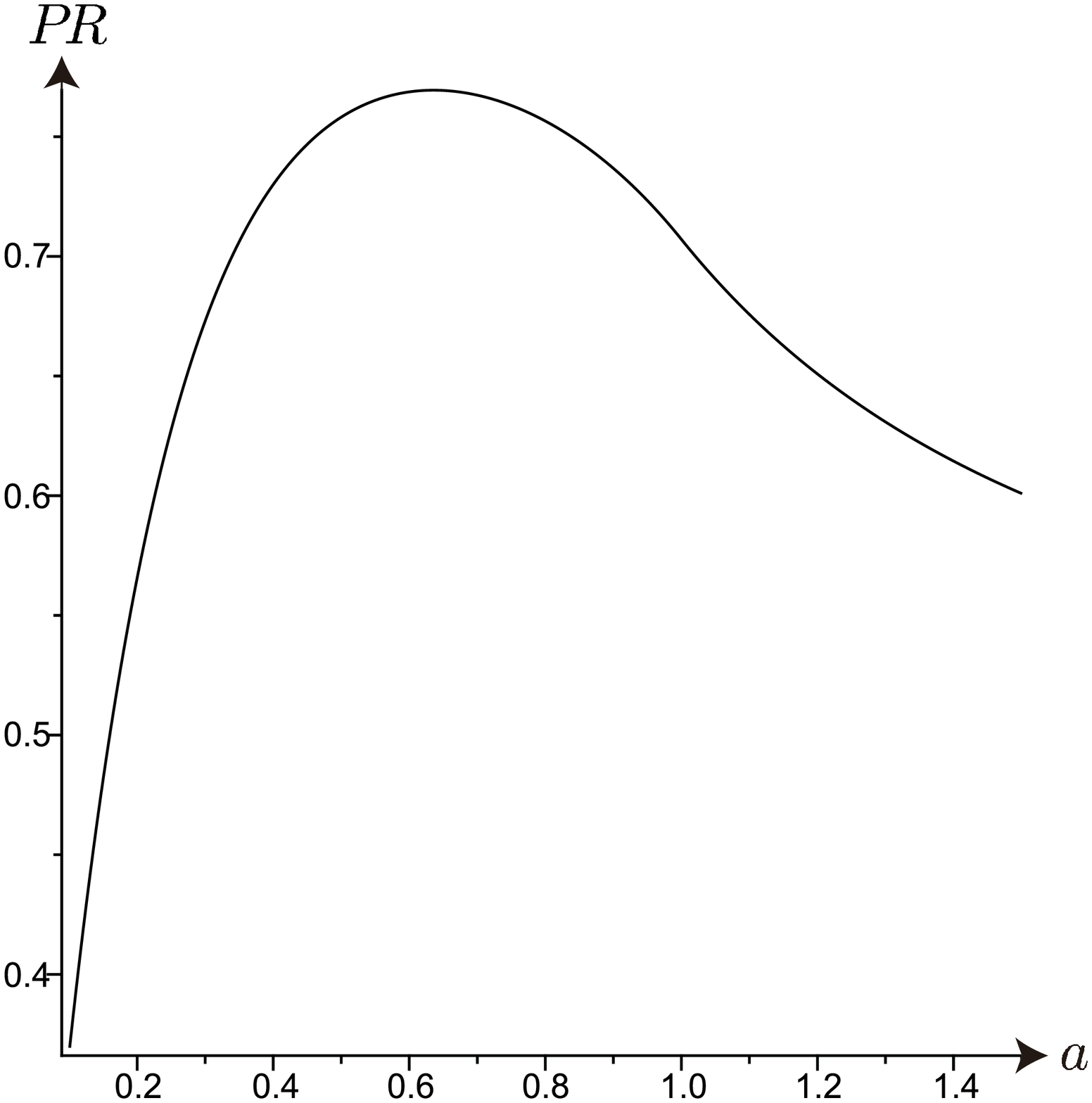}
\caption{The packing proportion of a double helix with slope $a$}
\label{pr}
\end{center}
\end{minipage}
\end{center}
\end{figure}

%
Next we compute the slope of ``best packed'' tubular double helix, namely, a double helix that maximizes the ratio of the volume of the fattest tubular neighbourhood and the volume of the circumscribed cylinder of the tubes (Figure \ref{0635805_cylinder}). 
This ``packing proportion function'' is given by 
$$
PR(a)=\frac{\pi \, \rho^2(a)\cdot AL(a)}{\pi(1+\rho(a))^2\cdot 2\pi a}
=\frac{2 \sqrt{1+a^2} \, \rho^2(a)}{a(1+\rho(a))^2},
$$
where $\rho(a)$ is the thickness of $\G(a)$. 
When $a\ge1$ we have $PR(a)=\sqrt{1+a^2}/2a$ which takes the maximum value at $\sqrt{2}/2\approx0.707107$ at $a=1$. The numerical computation using Maple implies that the packing proportion takes the maximum which is approximately equal to $0.7694$ at $a\approx 0.635805$ (Figure \ref{pr}), when the thickness $\rho$ is $0.832589$ and the ratio of the pitch and the thickness is $P/\rho\approx4.7981$. 

\begin{figure}[htbp]
\begin{center}
\begin{minipage}{.55\linewidth}
\begin{center}
\includegraphics[width=.75\linewidth]{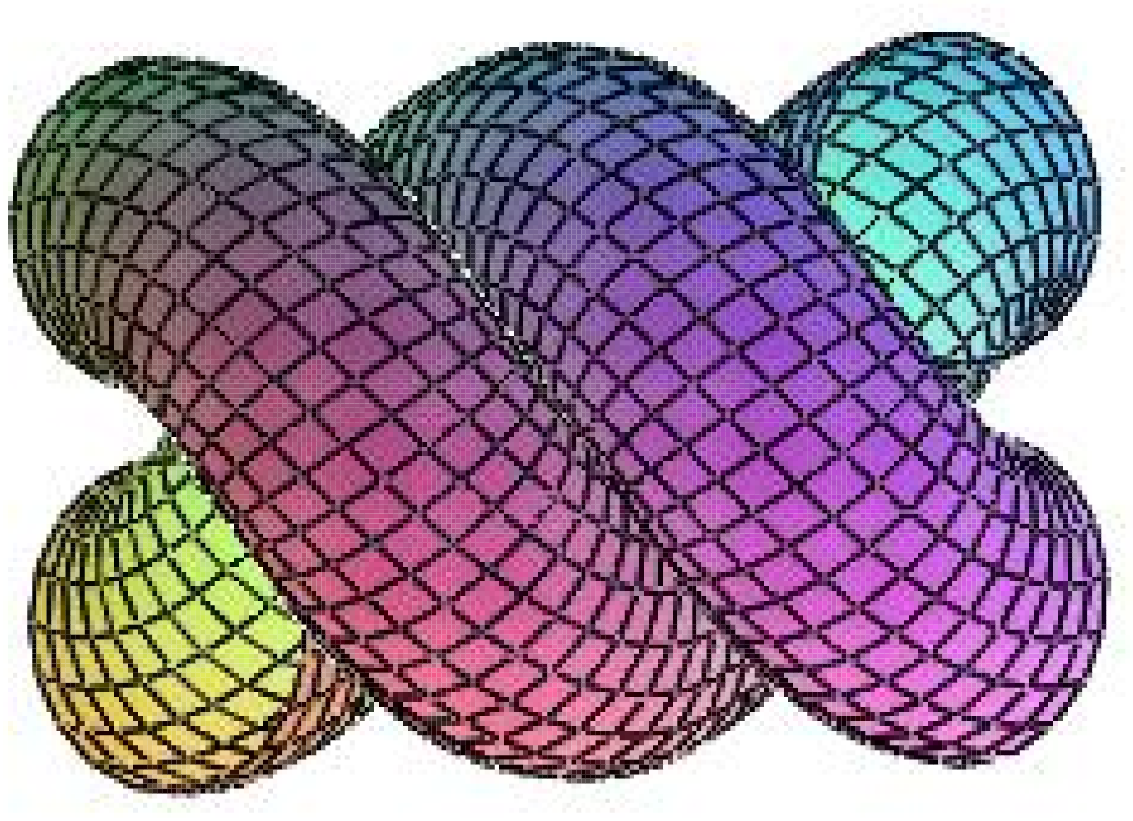}
\caption{Best volume packing tubular double helix}
\label{0635805}
\end{center}
\end{minipage}
\hskip 0.2cm
\begin{minipage}{.4\linewidth}
\begin{center}
\includegraphics[width=.75\linewidth]{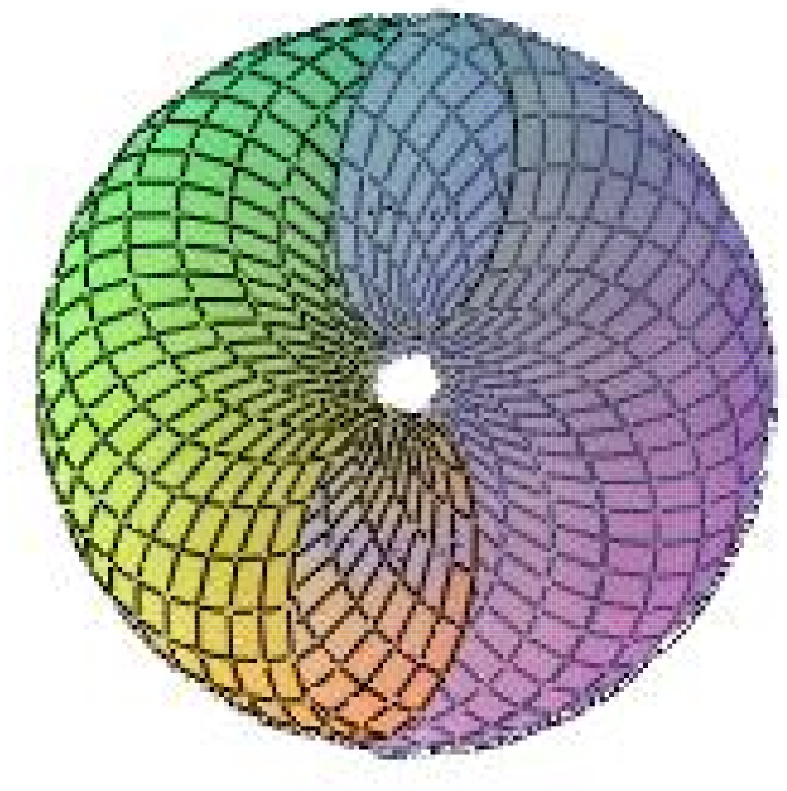}
\caption{The same one seen from an axial direction}
\label{0635805_axis}
\end{center}
\end{minipage}
\end{center}
\end{figure}

The best volume packing double helix had already been studied by Kasper Olsen and Jakob Bohr\cite{OB}. 

\section{Energy minimizing double helix}
\subsection{$r^{-\alpha}$-modified potential energy}
\subsubsection{Mutual energy}
Define the {\em $r^{-\alpha}$-modified Coulomb's potential} of $\G_2(a)$ at a point  $x_0$ in $\G_1(a)$ by 
$$
V^{(\alpha)}(a)=\int_{\G_2(a)}\frac{dy}{|x_0-y|^\alpha}
=\int_{-\infty}^\infty\frac{\sqrt{1+a^2}\,\,d\theta}
{{\left\{(\cos\theta+1)^2+\sin^2\theta+a^2\theta^2\right\}}^{\frac\alpha2}}\,.
$$
We assume $\alpha>1$ so that $V^{(\alpha)}(a)$ is finite. 
As $\alpha$ increases $V^{(\alpha)}(a)$ is dominated by the contribution of near-by points. 
Remark that it does not depend on the point $x_0\in\G_1(a)$ because of the symmetry. 
Define the {\em average cross-term $r^{-\alpha}$-modified Coulomb's potential energy} of $\G(a)$ by 
$$
AE^{(\alpha)}(a)=\frac12 AL(a)\cdot V^{(\alpha)}(a)
=2\pi(1+a^2)\int_{-\infty}^\infty\frac{d\theta}{{\left\{(\cos\theta+1)^2+\sin^2\theta+a^2\theta^2\right\}}^{\frac\alpha2}}\,.
$$
The asymptotic behavior of $AE^{(\alpha)}(a)$ as $a$ goes to $\infty$ or $0$ is given as follows: 
$AE^{(\alpha)}(a)\sim C_\infty \cdot a$ as $a\to+\infty$ and $AE^{(\alpha)}(a)\ge C_0 \cdot \frac1{a^{\alpha-1}}$ as $a\to+0$, where 
$C_\infty=2^{2-\alpha}\pi\int_{-\infty}^\infty(1+t^2)^{-\frac\alpha2}\,dt$ and $C_0=2\pi^{2-\alpha}\int_{-\infty}^\infty(1+t^2)^{-\frac\alpha2}\,dt$.

\begin{figure}[htbp]
\begin{center}
\begin{minipage}{.45\linewidth}
\begin{center}
\includegraphics[width=\linewidth]{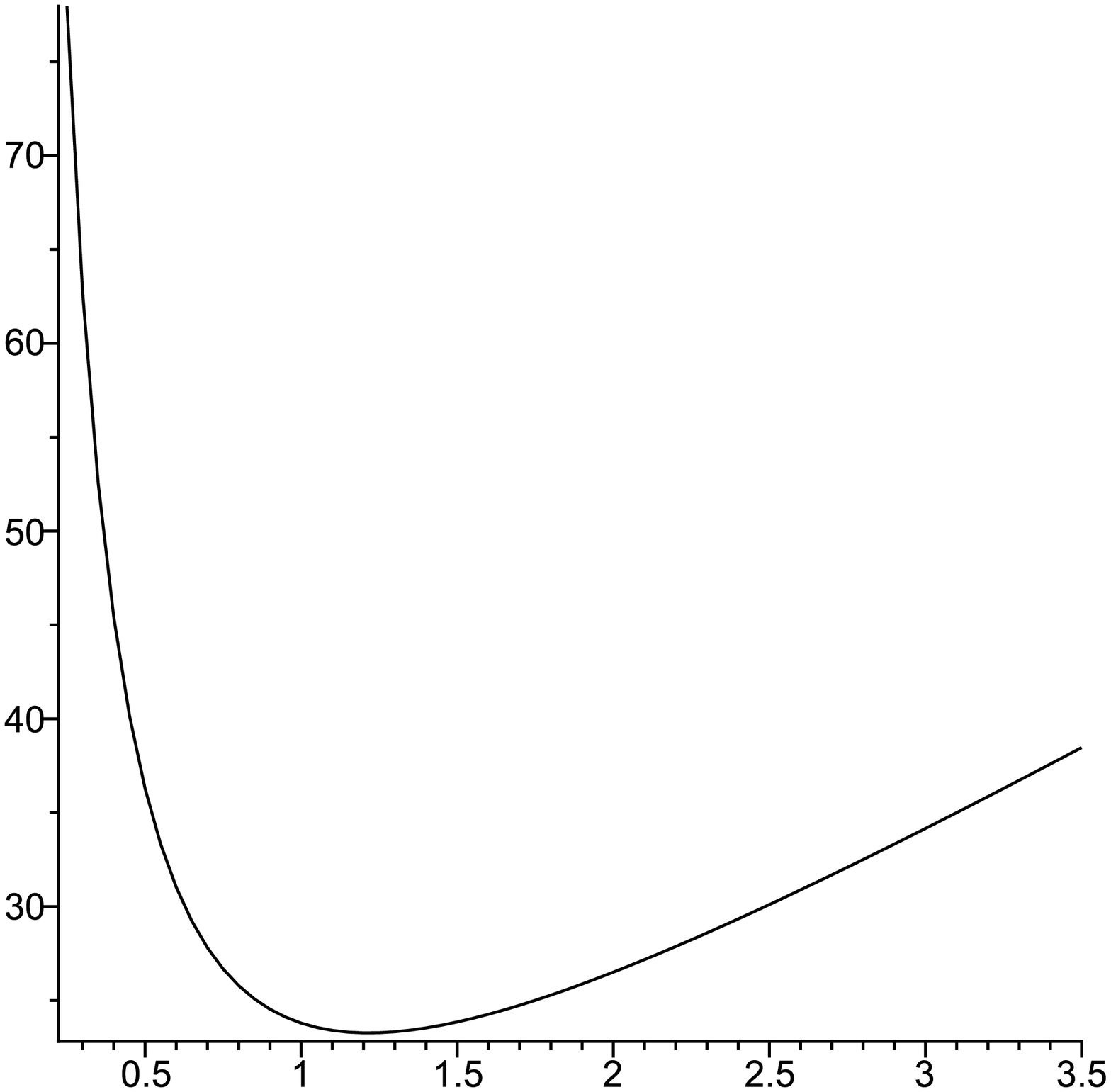}
\caption{$AE^{(2)}(a)$}
\label{m2-1}
\end{center}
\end{minipage}
\hskip 0.4cm
\begin{minipage}{.45\linewidth}
\begin{center}
\includegraphics[width=\linewidth]{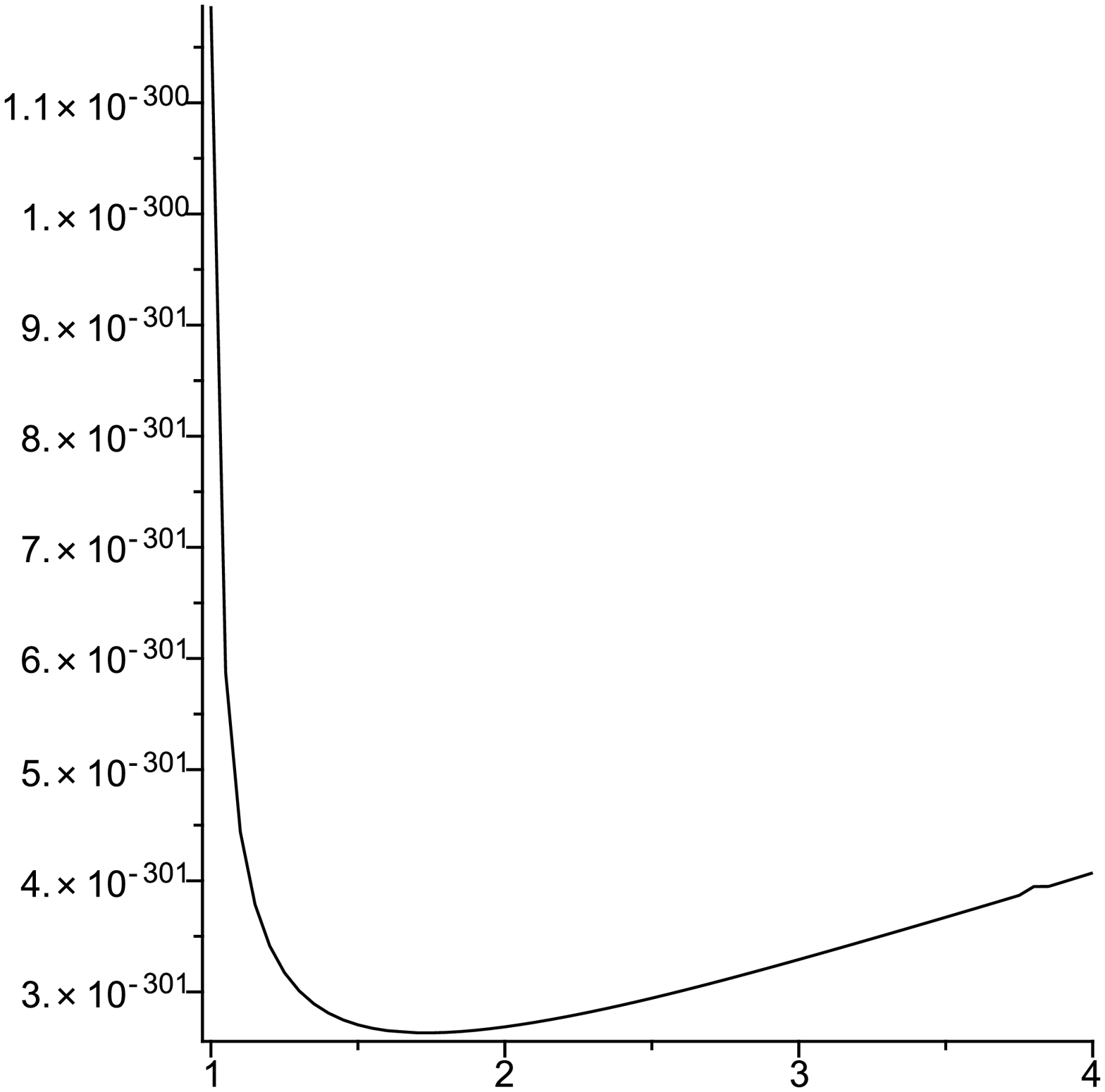}
\caption{$AE^{(1000)}(a)$}
\label{m1000-1}
\end{center}
\end{minipage}
\end{center}
\end{figure}

%
%
%
%
%
Numerical experiments using Maple imply that, for each $\alpha>1$, $AE^{(\alpha)}(a)$ is a convex function of $a$. The graphs of $AE^{(\alpha)}(a)$ when $\alpha=2$ and $1000$ are illustrated in Figures \ref{m2-1} and \ref{m1000-1}. 
%
%
\begin{figure}[htbp]
\begin{center}
\begin{minipage}{.47\linewidth}
\begin{center}
%
\par\noindent
\begin{tabular}{|c|c|c|}
\hline
$\alpha$ & $\stackrel{\phantom{.}}{a(\alpha)}$ & $P/\rho$ \\[1mm] 
\hline
\hline
$1.05$ & $1.01015 \sim 1.01018$ & 6.3471\\ \hline
$1.1$ & $1.02125 \sim 1.02130$ & 6.4169 \\ \hline 
$1.2$ & $1.04305 \sim 1.04311$ & 6.5539\\ \hline
$1.5$ & $1.109545 \sim 1.109555$ & 6.9715\\ \hline
$1.75$ & $1.16380 \sim 1.16382$ & 7.3124\\ \hline
$2$ & $1.21513 \sim 1.21515$ & 7.6349\\ \hline
$2.5$ & $1.30566 \sim 1.30568$ & 8.2038\\ \hline
$3$ & $1.378780 \sim 1.378786$ & 8.6631\\ \hline
$4$ & $1.48104 \sim 1.48112$ & 9.3059\\ \hline
$5$ & $1.54403 \sim 1.54406$ & 9.7015\\ \hline
$6$ & $1.58435 \sim 1.58445$ & 9.9551\\ \hline
$8$ & $1.63085 \sim 1.63095$ & 10.247 \\ \hline
$10$ & $1.65596 \sim 1.65598$ & 10.405\\ \hline
$15$ & $1.68555 \sim 1.68565$ & 10.591\\ \hline
$20$ & $1.69878 \sim 1.69881$ & 10.674\\ \hline
$40$ & $1.71655 \sim 1.71660$ & 10.786\\ \hline
$100$ & $1.72606 \sim 1.72617$ & 10.846\\ \hline
$150$ & $1.72810 \sim 1.72818$ & 10.858\\ \hline
$200$ & $1.72910 \sim 1.72913$ & 10.865\\ \hline
$400$ & $1.73056 \sim 1.73062$ & 10.874\\ \hline
$1000$ & $1.73139 \sim 1.73140$ & 10.878\\ \hline
\end{tabular}
\\
\bigskip
Table I. $AE^{(\alpha)}$-minimizing slopes $a(\alpha)$. 
The thickness is always $1$ as $a\ge1$. 
%
%
\end{center}
\end{minipage}
\hskip 0.2cm
\begin{minipage}{.47\linewidth}
\begin{center}
\includegraphics[width=\linewidth]{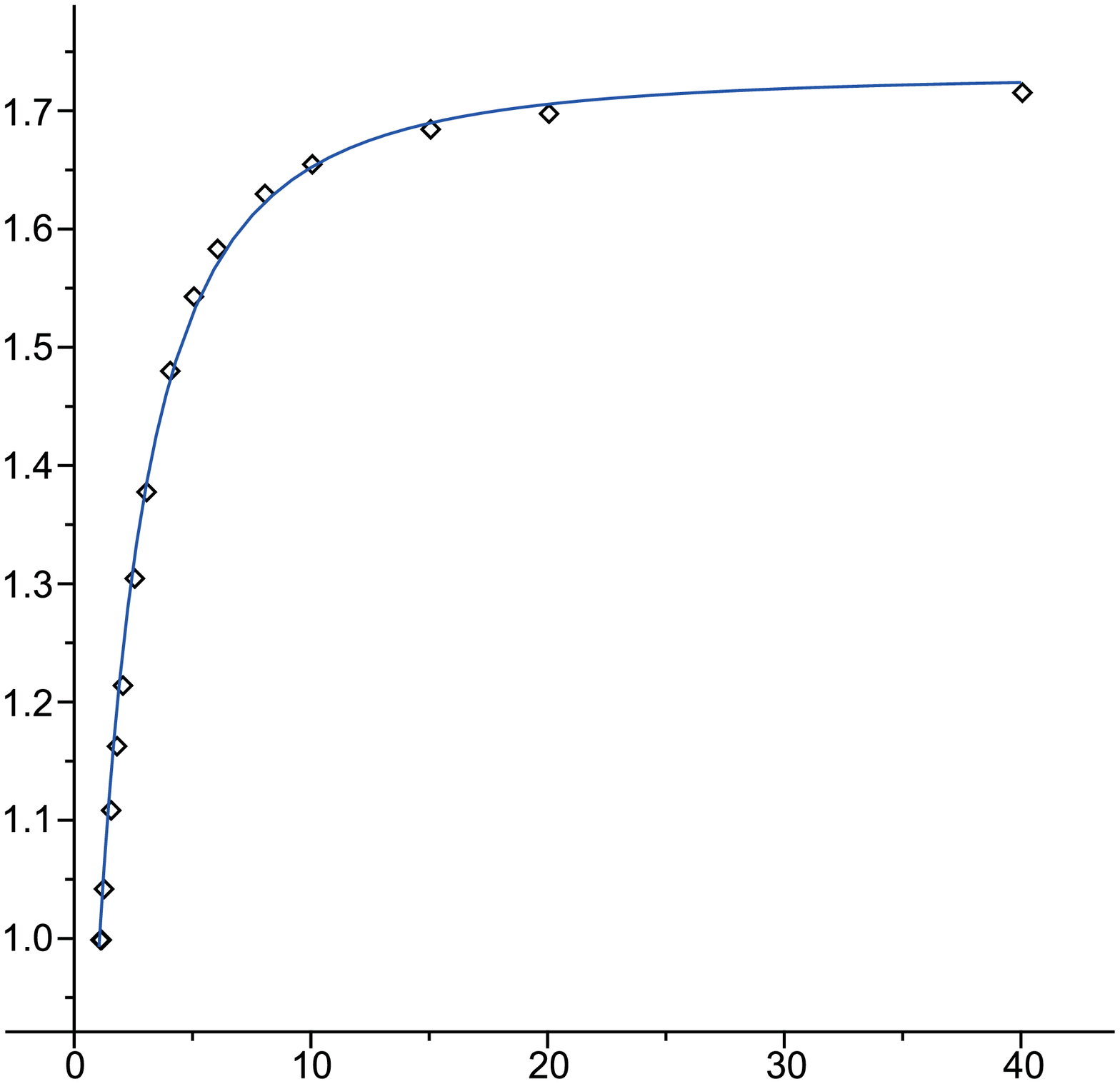}
\caption{The data of $a(\alpha)$ (diamond) and a fitting curve $f(\alpha)$ (blue)}
\label{fit_curve_energy_min_slopes}
\end{center}
\end{minipage}
\end{center}
\end{figure}

Let $a(\alpha)$ denote the slope that minimizes $AE^{(\alpha)}$. 
%
Their values obtained by numerical computations using Maple are shown in Table I. 
They are illustrated in dotted diamonds in Figure \ref{fit_curve_energy_min_slopes}. 
The computations when $\alpha$ equals $1.5, 200, 400$, and $1000$ were first done by Miyuki Tani. 
One of the candidates for a fitting curve which is drawn by a blue curve in Figure \ref{fit_curve_energy_min_slopes} is given by 
$$
f(\alpha)=1.73144-\frac{14.0806074006923936}{(\alpha+3.31656608512330786)^2}\,.
$$

\begin{figure}[htbp]
\begin{center}
\begin{minipage}{.28\linewidth}
\begin{center}
\includegraphics[width=\linewidth]{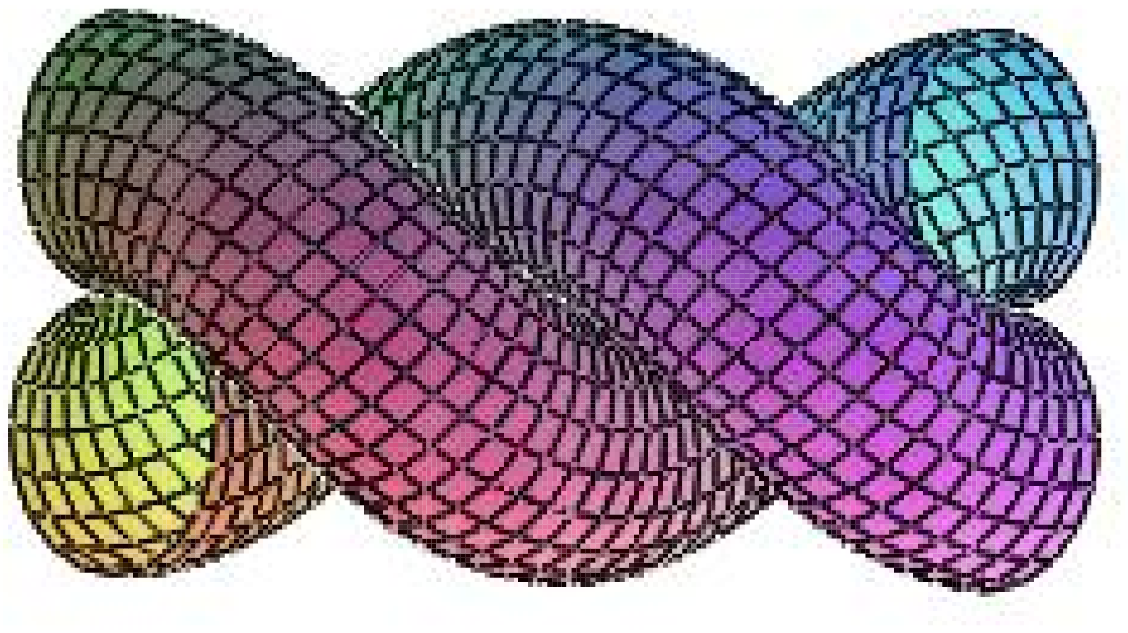}
\end{center}
\end{minipage}
\hskip 0.1cm
\begin{minipage}{.32\linewidth}
\begin{center}
\includegraphics[width=\linewidth]{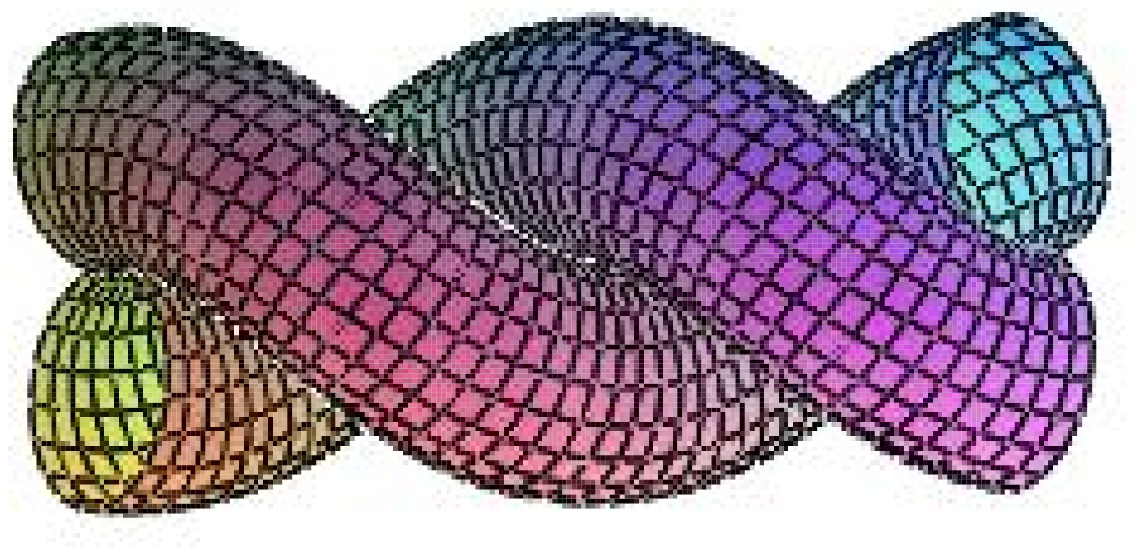}
\end{center}
\end{minipage}
\hskip 0.1cm
\begin{minipage}{.36\linewidth}
\begin{center}
\includegraphics[width=\linewidth]{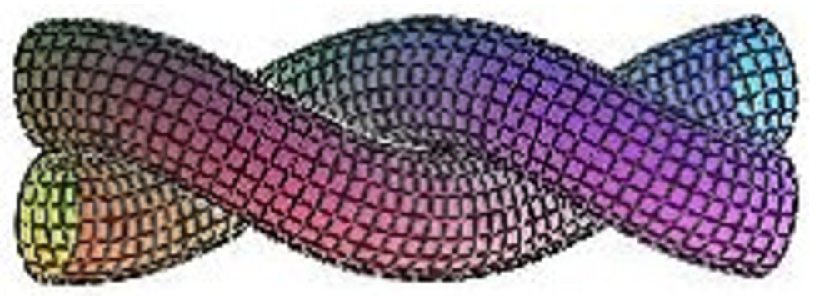}
\end{center}
\end{minipage}
\end{center}
\caption{Fattest tubes around an energy minimizing double helix $\G(a)$. (left) $a=1$ i.e. $P/\rho=2\pi$. We conjecture that it is the limit of $AE^{(\alpha)}$-minimizer as $\alpha$ goes down to $1$, and at the same time, the limit of $E^{\langle k\rangle}_{\exp}$-minimizer as $k$ goes down to $0$. (center) $a=1.21514$ i.e. $P/\rho=7.6349$. Optimal with respect to $AE^{(\alpha)}$ when $\alpha=2$. (right) $a=1.7314$ i.e. $P/\rho=10.878$. We conjecture that it is the limit of $AE^{(\alpha)}$-minimizer as $\alpha$ goes to $+\infty$, and at the same time, the limit of $E^{\langle k\rangle}_{\exp}$-minimizer as $k$ goes to $+\infty$. }
\end{figure}

\subsubsection{Total energy when $\alpha=2$}
The energy $AE^{(\alpha)}(a)$ is based on the interaction of two different strands $\G_1(a)$ and $\G_2(a)$. If we further take into account the interaction between the same strand, we obtain the ``{\sl total energy}''. As the integral diverges at the diagonal set, we need renormalization to get a well-defined functional. 

Define the {\em renormalized $r^{-2}$-modified Coulomb's potential} at a point $x_0$ in $\G_1(a)$ by\cite{O} 
$$
V^{(2)}_{\textrm{total}}(a)=\lim_{\varepsilon\to+0}\left(\int_{\G_1(a)\setminus N_\e(x_0)}\frac{dy}{|x_0-y|^2}-\frac2{\e}\right)+\int_{\G_2(a)}\frac{dz}{|x_0-z|^2}\,,
$$
where $N_\e(x_0)$ denotes the $\e$-neighbourhood of $x_0$ in $\G_1(a)$,  $\{y\in\G_1(a)\,:\,d(x_0,y)<\e\}$. Here $d(x_0,y)$ is the distance between $x_0$ and $y$, either the arc-length or $|x_0-y|$, both giving the same result\cite{O2}. Define the {\em average renormalized $r^{-2}$-modified Coulomb's potential energy} of $\G(a)$ per one twist by 
$$AE^{(2)}_{\textrm{total}}(a)=\frac12 AL(a)\cdot V^{(2)}_{\textrm{total}}(a)=2\pi\sqrt{1+a^2}\, V^{(2)}_{\textrm{total}}(a).$$ 
Rob Kusner and John Sullivan showed that it takes the minimum value at $a\approx 1.454$ \cite{Ku-Su1'}. Remark that it is greater than the value of the optimal slope for $AE^{(2)}$. This is because the ``self energy'' of $\G_1(a)$ is a decreasing function of $a$. 

\subsection{$\frac1r\exp(-kr)$-energy}\label{subs_e_exp}
Let $x_0$ be a point in $\G_1(a)$. Define 
$$
V^{\langle k\rangle}_{\exp}(a)=\int_{\G_2(a)}\frac1r\exp(-kr)\,dy
=2\int_0^\infty\frac{\exp\big(-k\sqrt{2+2\cos\theta+a^2\theta^2}\,\big)}{\sqrt{2+2\cos\theta+a^2\theta^2}}\sqrt{1+a^2}\,d\theta,
$$
where $r=|x_0-y|$. We assume $k>0$ so that $V^{\langle k\rangle}_{\exp}(a)$ is finite. As $k$ increases $V^{\langle k\rangle}_{\exp}(a)$ is dominated by the contribution of near-by points. 
Remark that it does not depend on the point $x_0\in\G_1(a)$ because of the symmetry.  
Define 
$$
E^{\langle k\rangle}_{\exp}(a)=\frac12 AL(a)\cdot V^{\langle k\rangle}_{\exp}(a)
=4\pi(1+a^2)\int_0^\infty\frac{\exp\big(-k\sqrt{2+2\cos\theta+a^2\theta^2}\,\big)}{\sqrt{2+2\cos\theta+a^2\theta^2}}\,d\theta.
$$

\begin{figure}[htbp]
\begin{center}
\begin{minipage}{.45\linewidth}
\begin{center}
\includegraphics[width=\linewidth]{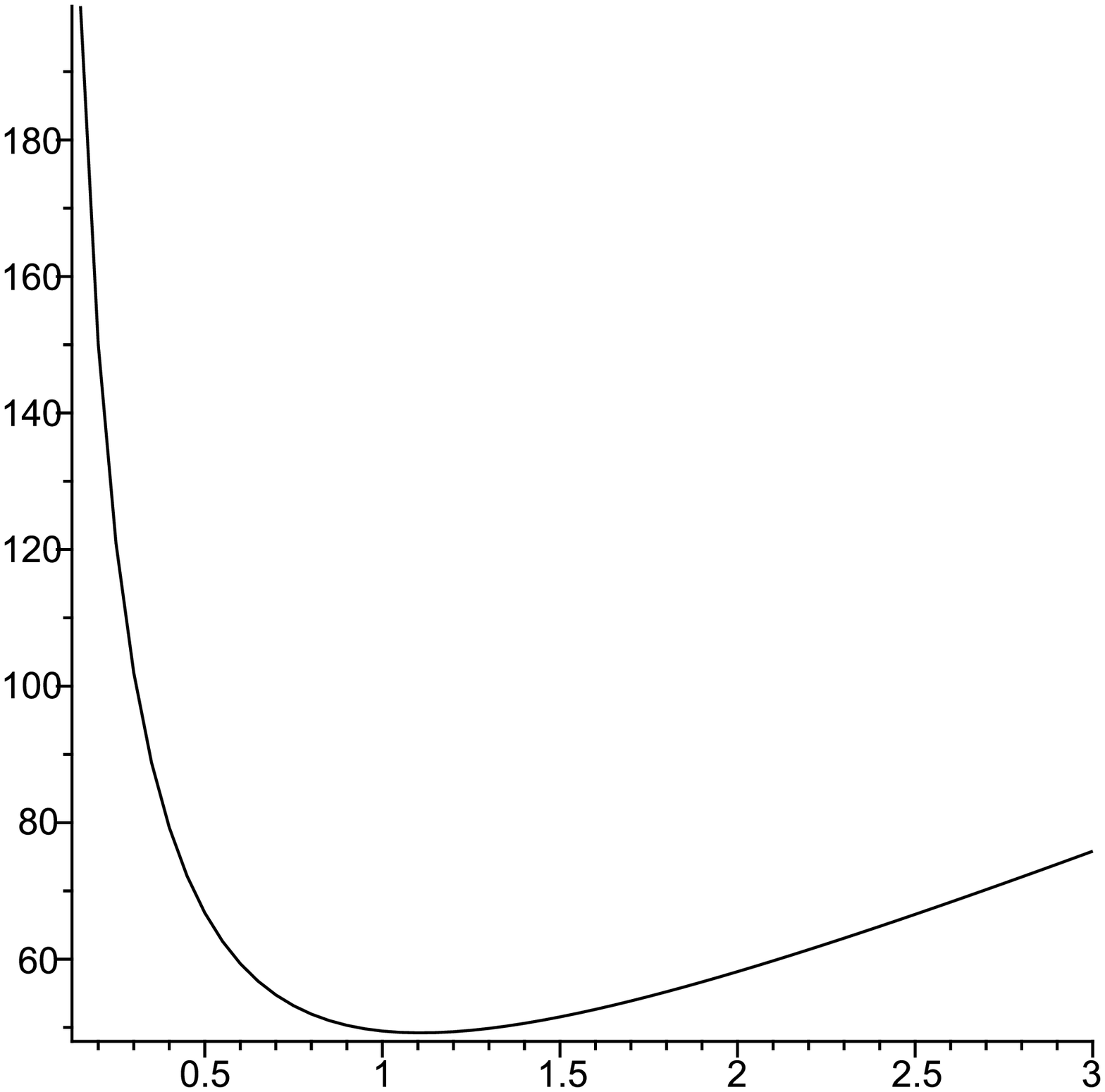}
\caption{$E^{\langle 0.1\rangle}_{\exp}(a)$}
\label{exp_energy_0p1}
\end{center}
\end{minipage}
\hskip 0.4cm
\begin{minipage}{.45\linewidth}
\begin{center}
\includegraphics[width=\linewidth]{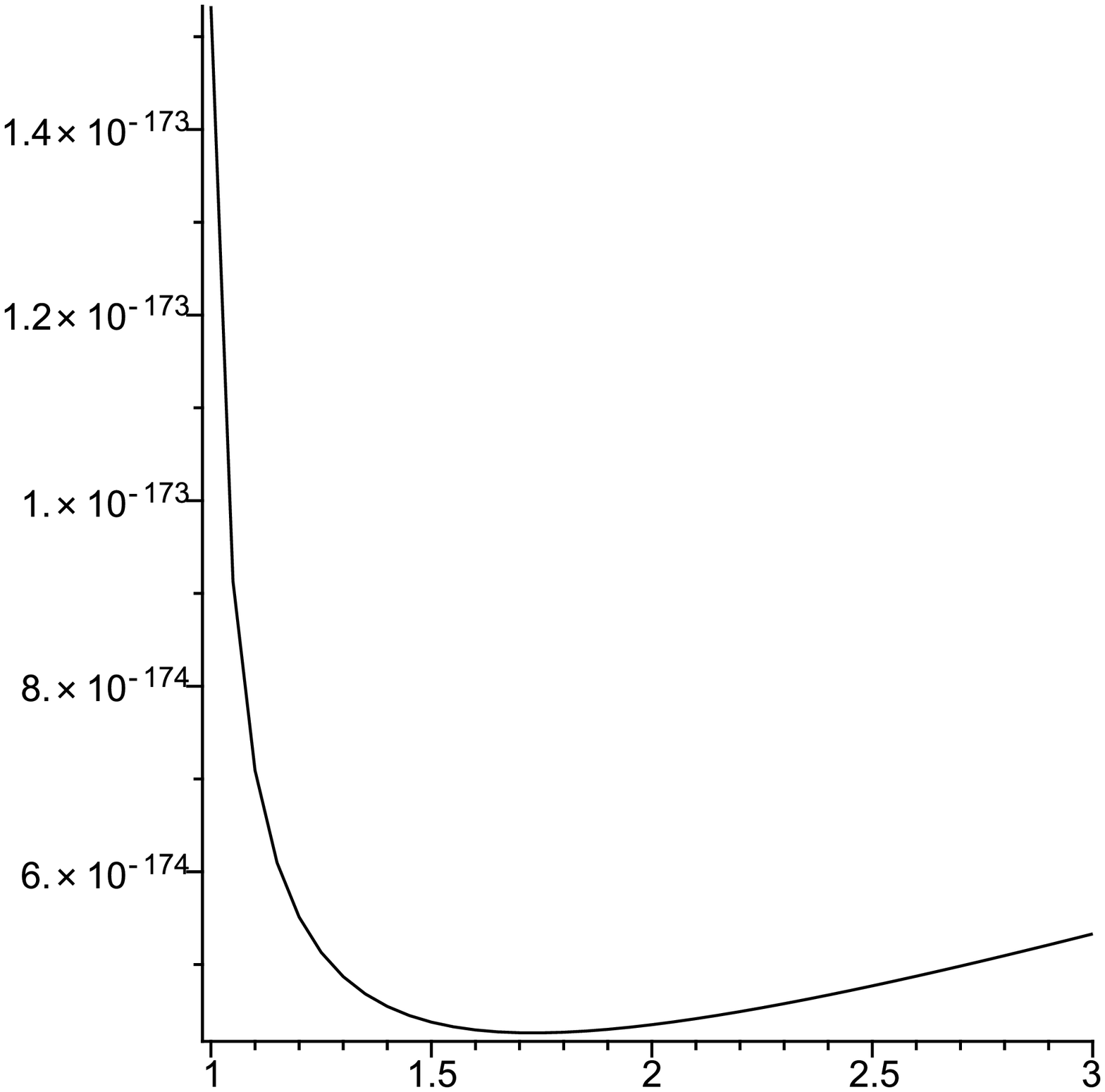}
\caption{$E^{\langle 200\rangle}_{\exp}(a)$}
\label{exp_energy_200}
\end{center}
\end{minipage}
\end{center}
\end{figure}

Numerical experiments using Maple imply that, for each $k>0$, $E^{\langle k\rangle}_{\exp}(a)$ is a convex function of $a$. The graphs of $E^{\langle k\rangle}_{\exp}(a)$ when $k=0.1$ and $200$ are illustrated in Figures \ref{exp_energy_0p1} and \ref{exp_energy_200}. 

\begin{figure}[htbp]
\begin{center}
\begin{minipage}{.49\linewidth}
\begin{center}
\begin{tabular}{|c|c|c|}
\hline
$k$ & $\stackrel{\phantom{.}}{a_{\exp}\langle k\rangle}$ & $P/\rho$ \\[1mm] 
\hline
\hline
$0.01$ & $1.01947$ & 6.4055\\ \hline
$0.1$ & $1.10680$ & 6.9542\\ \hline
$0.3$ & $1.20560$ & 7.5750\\ \hline
$0.5$ & $1.29850$ & 8.1587\\ \hline
$0.75$ & $1.39250$ & 8.7493\\ \hline
$1$ & $1.45953$ & 9.1705\\ \hline
$1.5$ & $1.5409$ & 9.6818\\ \hline
$2$ & $1.58628$ & 9.9669\\ \hline
$3$ & $1.6340$ & 10.267\\ \hline
$4$ & $1.65845$ & 10.420\\ \hline
$5$ & $1.6732$ & 10.513\\ \hline
$6$ & $1.6831$ & 10.575\\ \hline
$8$ & $1.69543$ & 10.653\\ \hline
$10$ & $1.70282$ & 10.699\\ \hline
$20$ & $1.7175$ & 10.791\\ \hline
$30$ & $1.72238$ & 10.822\\ \hline
$100$ & $1.72916$ & 10.865\\ \hline
$200$ & $1.7306$ & 10.874\\ \hline
\end{tabular}
\\
\bigskip
Table II. $E^{\langle k\rangle}_{\exp}$-minimizing slopes $a_{\exp}\langle k\rangle$. 
The thickness is always $1$ as $a\ge1$. 
%
\end{center}
\end{minipage}
\hskip 0.1cm
\begin{minipage}{.47\linewidth}
\begin{center}
\includegraphics[width=\linewidth]{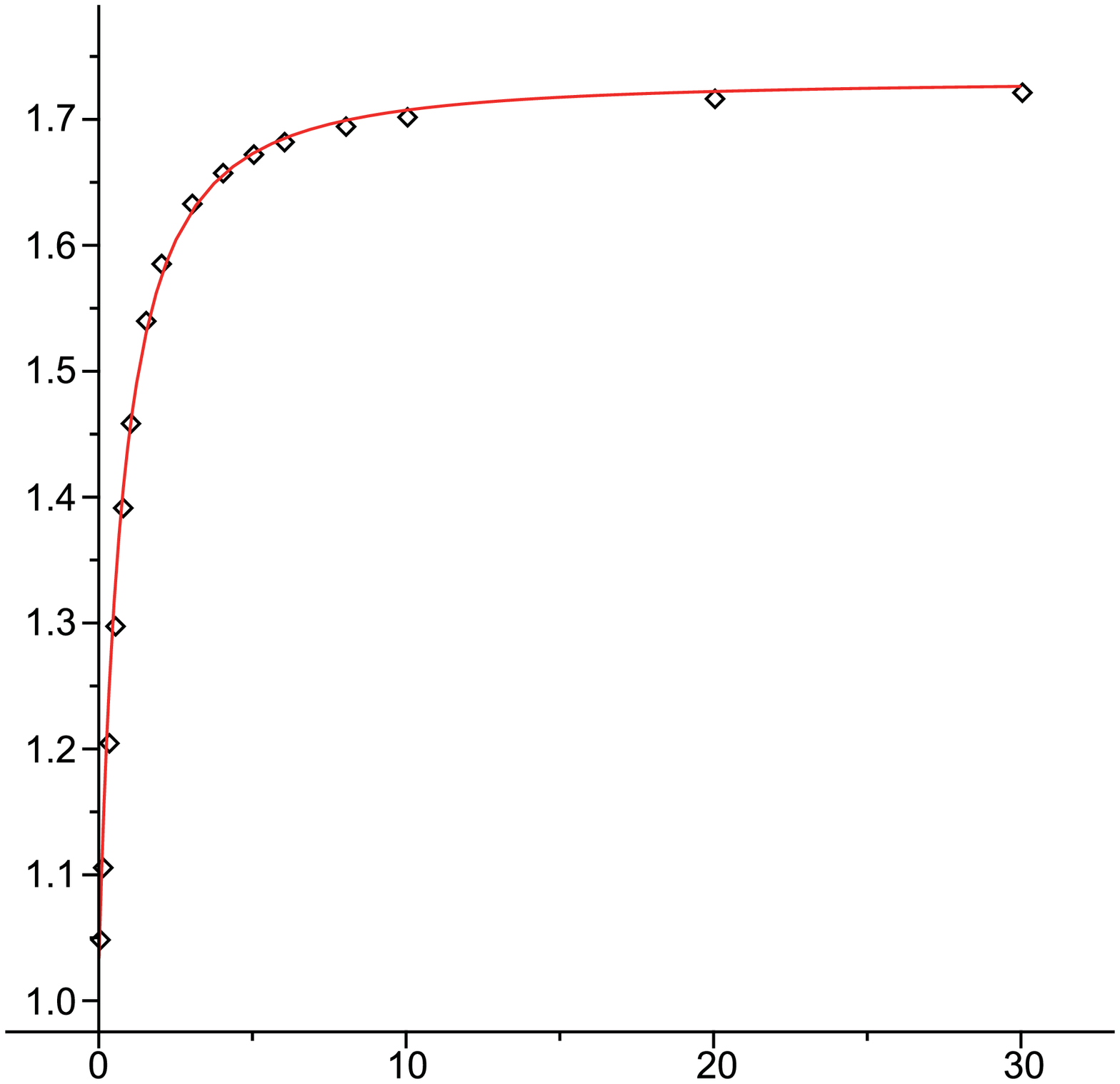}
\caption{The data of $a_{\exp}\langle k\rangle$ (diamond) and a fitting curve $g(k)$ (red)}
\label{fit_energy_exp}
\end{center}
\end{minipage}
\end{center}
\end{figure}

Let $a_{\exp}\langle k\rangle$ denote the slope that minimizes $E^{\langle k\rangle}_{\exp}$. Their values obtained by numerical computations using Maple are shown in Table II. They are illustrated in dotted diamonds in Figure \ref{fit_energy_exp}. One of the candidates for a fitting curve which is drawn by a red curve in Figure \ref{fit_energy_exp} is given by  
$$
g(k)=1.7314-\frac{0.899179433232235592}{(k+1.17422998242145948)^\frac32}\,.
$$

\subsection{Conjectures}

\bigskip
We like to end the section with conjectures implied by numerical experiments. 

\medskip
\noindent
{\bf Conjecture}. Let $\alpha>1$ and $k>0$. 

\begin{enumerate}
\item For fixed $\alpha$, $AE^{(\alpha)}(a)$ is a convex function of $a$. 

For fixed $k$, $E^{\langle k\rangle}_{\exp}(a)$ is a convex function of $a$. 
\item The slope $a(\alpha)$ that minimizes $AE^{(\alpha)}(a)$ is an increasing function of $\alpha$. 

The slope $a_{\exp}\langle k\rangle$ that minimizes $E^{\langle k\rangle}_{\exp}$ is an increasing function of $k$. 
\item Both $\{a(\alpha)\,|\,\alpha>1\}$ and $\{a_{\exp}\langle k\rangle\,|\,k>0\}$ are bounded subsets of $\mathbb R$, and we have $\displaystyle \lim_{\alpha\to1+0}a(\alpha)=\lim_{k\to+0}a_{\exp}\langle k\rangle=1$ and $\displaystyle \lim_{\alpha\to+\infty}a(\alpha)=\lim_{k\to+\infty}a_{\exp}\langle k\rangle\le\sqrt 3$. 
\end{enumerate}

\section{Conclusion}
We have studied three kinds of functionals for double helices with straight axes and compute the optimal slopes with respect to them numerically (i.e. without mathematical proofs). 

\begin{figure}[htbp]
\begin{center}
\includegraphics[width=.5\linewidth]{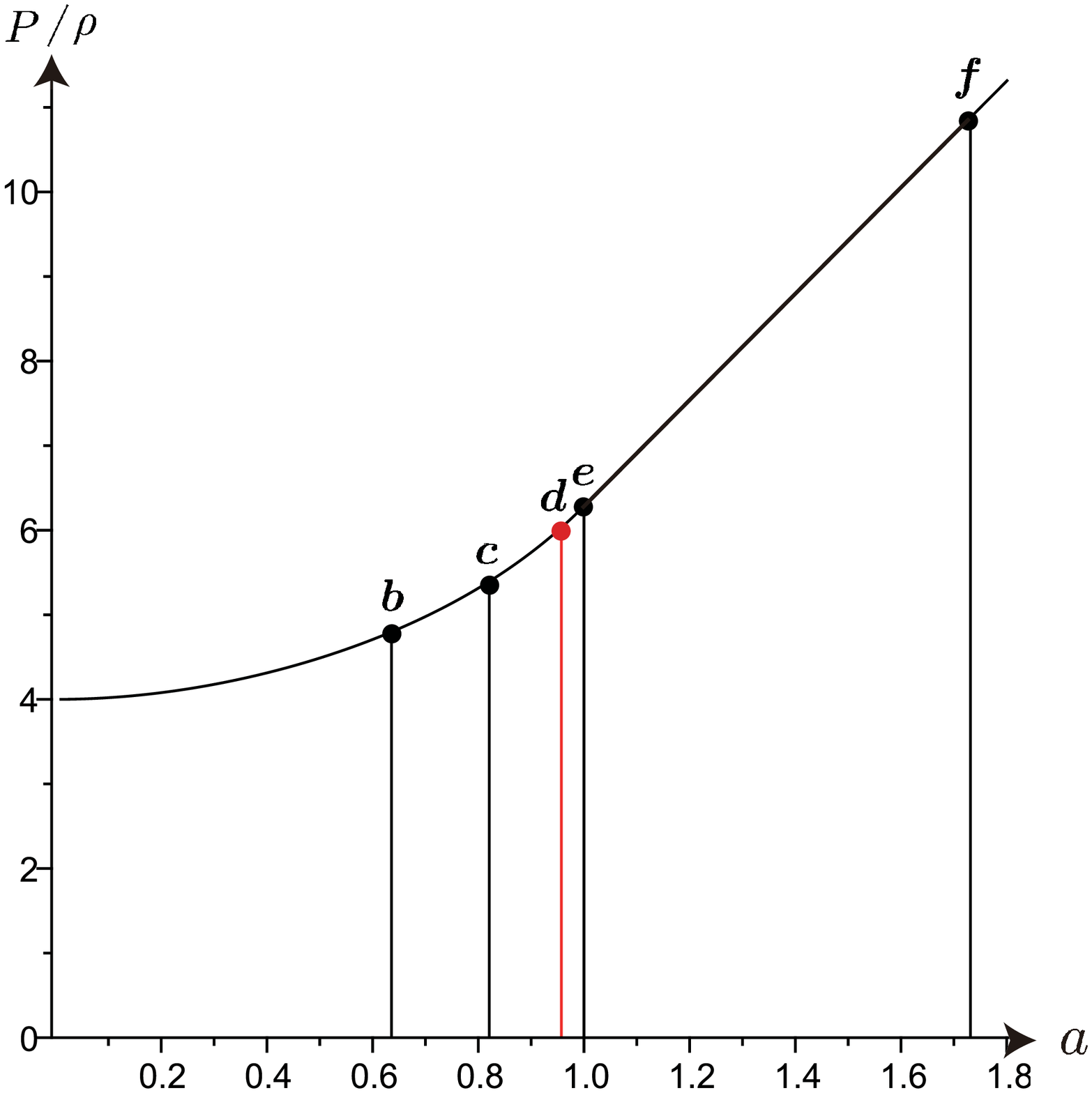}
\caption{The slope $a$ and the ratio of the pitch and the thickness $P/\rho$, and the optimal ones with respect to various functionals: ({\bf \it b}\,) volume packing ratio ({\bf \it c}\,) ropelength (ideal double helix) ({\bf \it d}\,) DNA ({\bf \it e}\,) the limit of the interaction energy $AE^{(\alpha)}$ as $\alpha\searrow1$ or $E^{\langle k\rangle}_{\exp}$ as $k\searrow0$ (conjecture) ({\bf \it f}\,) the limit of the interaction energy $AE^{(\alpha)}$ as $\alpha\nearrow+\infty$ or $E^{\langle k\rangle}_{\exp}$ as $k\nearrow+\infty$ (conjecture). 
Any point between $e$ and $f$ can be realized as an optimal slope with respect to some interaction energy. }
\label{graph}
\end{center}
\end{figure}

If we consider the average ropelength, or equivalently, the ratio of the volume of the fattest tubular neighbourhood and the cube of the length, the maximum is attained by an {\sl ideal} double helix, whose slope is $0.82074$ (Figure \ref{graph}, {\bf \it c}). 

If we consider the ratio of the volume of the fattest tubular neighbourhood and that of a circumscribed cylinder of the tubes, the maximum is attained by a best volume packing double helix, whose slope is $0.635805$ (Figure \ref{graph}, {\bf \it b}). 

If we consider two kinds of one parameter families of interaction energies between two strands of types $r^{-\alpha}$ (modified Coulomb potential) and $\frac1r\exp(-kr)$ (screened Coulomb's potential or Yukawa potential), the optimal slopes seem to be between $1$ and $\sqrt3$ in both cases, approaching $1$ (or $1.73144$) (Figure \ref{graph}, {\bf \it e,f}\,) as the energy is dominated by the global (or respectively local) contribution, i.e. the contribution of near-by points gets weaker (or respectively stronger). 

Stasiak and Maddocks\cite{SM} reported that, in the case of DNA, the ratio $P/\rho$ is $6.03$ (Figure \ref{graph}, {\bf \it d}\,), which is within 4$\%$ error of that of optimal double helix if we use ``$r^{-1}$-potential energy'', i.e. the limit of either interaction energy as the parameter approaches to the critical value when the energy fails to be well-defined, when the contribution of near-by points gets weakest. 

\section*{Acknowledgements}
The author would like to thank the organizers and participants of the conference ``Statistical Physics and Topology of Polymers with Ramifications to Structure and Function of DNA and Proteins'' for stimulating conversations. 
In fact, he started to compute the energy $E^{\langle k\rangle}_{\exp}$ after a suggestion by a participant after the talk. The author is sorry not to have recognized the name. 
The author would also like to thank Andrzej Stasiak and the referee for helpful suggestions and the references, and Kasper Olsen and Jakob Bohr for new information and references.

\bigskip \noindent
Department of Mathematics and Information Sciences, \\
Tokyo Metropolitan University, \\
1-1 Minami-Ohsawa, Hachiouji-Shi, Tokyo 192-0397, JAPAN. \\
E-mail: ohara@tmu.ac.jp

\end{document}